\documentclass[onecolumn,nofootinbib,superscriptaddress,notitlepage,longbibliography,thightenlines,11pt]{revtex4-1}  

\usepackage[english]{babel}

\newcommand{\degree}{\ensuremath{^\circ}}

\usepackage[dvips]{graphicx}
\usepackage{booktabs}
\usepackage[table]{xcolor}
\usepackage{tabularx}
\usepackage{colortbl}
\usepackage{graphicx}
\usepackage{amsmath}

\usepackage{color}
\definecolor{darkblue}{rgb}{0,0,0.5}
\definecolor{lila}{rgb}{0.3,0,0.3}
\definecolor{turq}{rgb}{0,0.1,0.4}
\definecolor{lightblue}{rgb}{0.7,0.7,0.9}
\usepackage{url} 

\usepackage[pdftex,
 colorlinks=true,
  backref=page,
  linkcolor=darkblue, 
  filecolor=red,
  citecolor=turq, 
  urlcolor=lila, 
  pdftitle={Modification of the Absorption Cross Section in the Optical Near-field},
  pdfauthor={Moritz Striebel, Jeff F.Young, Jorg Wrachtrup, Ilja Gerhardt},
  pdfsubject={Scattering, Absorption, Extinction, Albedo, Light-Matter Interaction, Single Emitters},
  pdfkeywords={Scattering, Absorption, Extinction, Albedo, Light-Matter Interaction, Single Emitters},
  pdfpagelabels=true, 
  breaklinks=false,
  plainpages=false,
  backref=false,
  bookmarks,
  bookmarksnumbered=true]{hyperref}

\begin{document}

\title{Modification of the Absorption Cross Section in the Optical Near-field}

\author{Moritz Striebel}
\address{3. Physikalisches Institut, Universit\"at Stuttgart and Stuttgart Research Center of Photonic Engineering (SCoPE), Pfaffenwaldring 57, Stuttgart, D-70569, Germany}

\author{Jeff F. Young}
\address{University of British Columbia, 6224 Agricultural Rd, Vancouver, BC, V6T 1Z1, Canada}

\author{J\"org Wrachtrup}
\address{3. Physikalisches Institut, Universit\"at Stuttgart and Stuttgart Research Center of Photonic Engineering (SCoPE), Pfaffenwaldring 57, Stuttgart, D-70569, Germany}
\address{Max Planck Institute for Solid State Research, Heisenbergstra\ss e 1, D-70569 Stuttgart, Germany}

\author{Ilja Gerhardt}
\email{i.gerhardt@fkf.mpg.de}
\address{3. Physikalisches Institut, Universit\"at Stuttgart and Stuttgart Research Center of Photonic Engineering (SCoPE), Pfaffenwaldring 57, Stuttgart, D-70569, Germany}
\address{Max Planck Institute for Solid State Research, Heisenbergstra\ss e 1, D-70569 Stuttgart, Germany}

\begin{abstract}
The optical interaction of light and matter is modeled as an oscillating dipole in a plane wave. We analyze absorption, scattering and extinction for this system by the energy flow, which is depicted by streamlines of the Poynting vector. Depending on the dissipative damping of the oscillator, the streamlines end up in the dipole. Based on a graphical investigation of the streamlines, this represents the absorption cross section, and forms a far-field absorption aperture. In the near-field of the oscillator, a modification of the aperture is observed. This scheme can be adapted to a single dipolar emitter, interacting with a light field. In the case of the absorption by a single atom, where the oscillator has a circular dipole characteristics, we model the energy flow and derive the apertures.
\noindent 
\end{abstract}

\maketitle
\section{Introduction}

One of the most fundamental processes in light-matter interaction is the attenuation of light by matter. The introduced weakening of light is observed since ancient times. Today, in modern nano-optics, it is possible to perform controlled experiments with single entities, i.e.\ single photons~\cite{rezus_prl_2012} and single emitters~\cite{gerhardt_prl_2007}. The theoretical description of this optical interaction does not only address very fundamental questions, it also allows the optimization of the efficiency in light-matter interaction. The spectroscopy and optical research on single emitters in the past was engineered for optimal light-extraction strategies~\cite{barnes_epjd_2002,lee_np_2011}, such that e.g.\ a single molecule or a nano-crystal can be detected by fluorescence with a good signal to noise ratio. But it was rarely optimized to achieve the most efficient excitation. For optical protocols, such as the quantum phase gate, or coherent microscopy schemes~\cite{gerhardt_pra_2010}, an efficient interaction of a single photon to an external field, ideally a single photon, is desirable.

Experimentally, single molecules~\cite{plakhotnik_prl_2001,gerhardt_prl_2007,wrigge_np_2008}, quantum dots~\cite{vamivakas_nl_2007}, and atoms~\cite{wineland_ol_1987,tey_np_2008,tey_njp_2009,streed_absorption_2012} were researched in the past decade for their ability to attenuate a light beam. The solid state experiments were initially conducted under cryogenic conditions~\cite{gerhardt_prl_2007,wrigge_np_2008}. Only balanced detection made it possible to extend these fundamental studies to room temperature~\cite{kukura_nl_2008,celebrano_np_2011}. Motivated by the experimental realization of these fundamental processes, numerous theoretical approaches were reinvented~\cite{zumofen_prl_2008,zumofen_nc_2008}, often founded on electrodynamic calculations from the 1960s when the field of antenna theory was very active. One central question was the maximum amount of extinction in a real optical focus~\cite{zumofen_prl_2008}. Others describe the energy relations in the context of the optical theorem and in the near-field~\cite{davis_josaa_2001}. The energy flow itself, depicted as streamlines of the Poynting vector, introduced a vivid insight in this fundamental problem already decades ago~\cite{paul_spu_1983,bohren_ajop_1983}. Today, we see the most fundamental approaches are getting unified between classical antenna theory, quantum technologies and nano-optics.

In this paper we review the fundamental basics of coherent light-matter interaction and absorption of a single emitter. The relevant cross sections of absorption, extinction and scattering are analytically derived from first principles. Everything is extended to the mathematical description of the energy flow in the proximity of a dipolar emitter. While this was discussed already in the past decade, we address shortcomings and present novel findings. The cross sections are calculated, and the analytic result and their derivation by following the energy flow are compared. In the close proximity of the dipole the total field is altered by the presence of the emitter. This change also alters the effective cross sections. The (optical) near-field cross sections become more and more a dipolar shape. Whereas the dipolar emitter was assumed as a linear, Hertzian, dipole in many papers, we see the experimental efforts extended to single atoms~\cite{tey_np_2008}, which usually exhibit a circular dipole. Therefore, we adapt the underlying math to the case of a circular dipole.

\section{Absorption, Scattering and Extinction}

\begin{figure}[b]
  \includegraphics[width=\columnwidth]{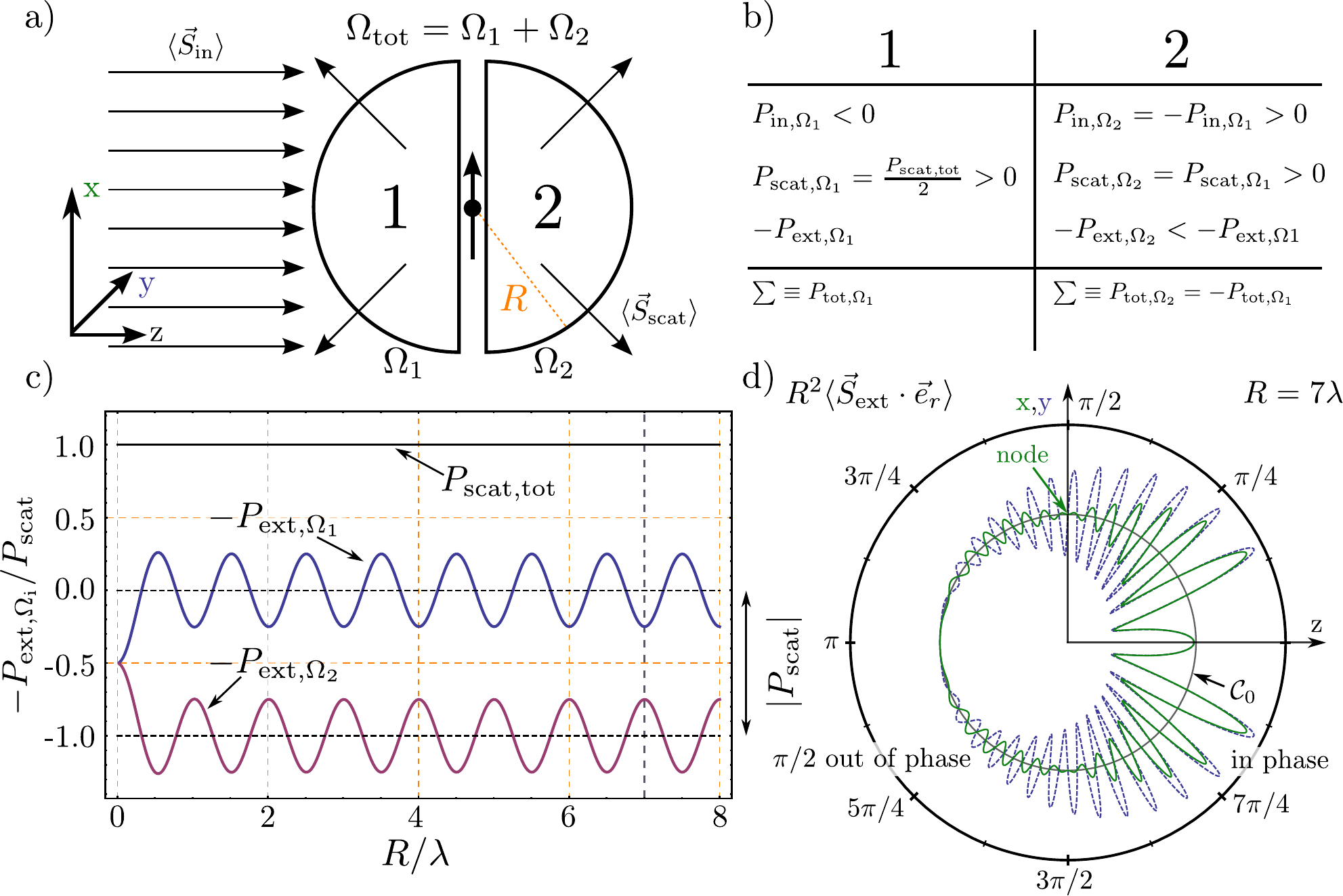}
  \caption{a) An electromagnetic wave is incident on a dipole, which scatters all energy without further loss. An integrating sphere is divided into a forward (2) and a backward (1) direction b) The different energy contributions and the corresponding signs i.e.\ if the energy exits (+) or enters the sphere (-). c) Forward ($-P_{\mathrm{ext,\Omega_2}}$,red) and backward ($P_{\mathrm{ext,\Omega_1}}$,blue) component dependent on the radius of the sphere as well as $P_{\mathrm{scat,tot}}$. The situation corresponding to d) is marked with a gray dashed line. d) Solutions for $\left. R^2 \langle \vec{S}_{\mathrm{ext}}\rangle \vec{e}_{\mathrm{r}} \right|_{\Theta=\pi/2}$ (blue, dashed) and $\left. R^2 \langle \vec{S}_{\mathrm{ext}}\rangle \vec{e}_{\mathrm{r}}\right|_{\phi =0}$ (green). The reference curve $\mathcal{C}_0$ defines zero. The magnitude of the solutions is plotted against the reference curve.}
  \label{fig:01}
\end{figure}

We approach the relevant math with an intuitive picture of the fundamental process of light-matter interaction. To that end we study the most basic case, given by the interaction of an incident plane wave with a single dipole. A single dipole emitter is assumed in free space at the origin of a coordinate system (Fig.~\ref{fig:01}a). An incident plane wave, propagating from $-z$ to $+z$ excites the emitter. This field is expressed as

\begin{equation}\label{formel2}
\vec{E}_{\mathrm{in}} = \vec{E}_{\mathrm{in}}^{\mathrm{0}} \exp(\mathrm{i} k z-\rm i \omega t) \;.
\end{equation}

\noindent
If we consider a electrically polarizable system -- such as a dipole -- in vacuum, excited by a incident electric field the response of the system due to the incident field is given by

\begin{equation}\label{formel1}
\vec{d} = \alpha \vec{E}_{\mathrm{in}} \;,
\end{equation}

\noindent
where $\vec{E}_{\mathrm{in}}$ is the electric field at the position of the emitter, $\alpha$ is the electric polarizability, and $\vec{d}$ is the induced dipole moment. We consider a parallel response of the system due to the incident field. The system is treated as a point-like dipole, i.e.\ infinitely small. In general, the response of the system is not necessarily parallel to the incident field, and $\alpha$ has to be written as a tensor. As a consequence of the excitation, the dipole scatters a certain amount of power into all $4\pi$ steradians of the environment (Fig.~\ref{fig:01}a).

In the following we calculate the time averaged energy flow into the system, to get an idea how energy is absorbed by a dipole. To do so, we calculate the time averaged Poynting vector, computed by (for time harmonic fields)~\cite{Jackson1975}

\begin{equation}\label{formel16}
\vec{S} = \frac{c}{8 \pi}\mathfrak{Re}\{\vec{E}_{\mathrm{tot}} + \vec{B}_{\mathrm{tot}}^*\} \;,
\end{equation} 

\noindent
where $\vec{E}_{\mathrm{tot}}$ is the total electric field calculated as a superposition of the incident field and the scattered field and $\vec{B}_{\mathrm{tot}}$ is the corresponding magnetic field. While the incident field is given by 

\begin{subequations}
\begin{eqnarray}\label{formel17}
\vec{E}_{\mathrm{in}} &=& E_{\mathrm{in}}^{0} \exp(\rm i kr \; \hat{r}\hat{n}_{\mathrm{in}}) \hat n_{\mathrm{in}}\\
\vec{B}_{\mathrm{in}} &=& \hat{k} \times \vec{E}_{\mathrm{in}}
\end{eqnarray}
\end{subequations}

\noindent
and the scattered dipole field is expressed by~\cite{Jackson1975} 

\begin{subequations}
\begin{eqnarray}\label{eqna:formel18} 
\vec{E}_{\mathrm{scat}} &=&d_0\left\{ \frac{k^2}{r} \left((\hat r \times \hat{d})\times \hat r \right)+ \left[3(\hat r(\hat r\cdot \hat{d}-\hat{d}))\right]\left(\frac{1}{r^3}-\frac{\mathrm i k}{r^2}\right) \right\} \exp{(\mathrm i k r)} \\
\vec{B}_{\mathrm{scat}}&=& k^2 d_0 (\hat r\times \hat{d})\left(1-\frac{1}{\mathrm i k r}\right) \frac{\exp{(\mathrm i k r)}}{r}\;.
\end{eqnarray}
\end{subequations}

\noindent
where $\hat n_{\mathrm{in}}$ is the unit vector in the direction of the incident polarizability, $\hat d$ is the unit vector in the direction of the induced dipole moment and $d_0$ is the magnitude of the dipole moment ($\vec{d} = d_0 \cdot \hat{d}$). Since we said that the dipole moment is parallel to the incident wave, $\hat n_{\mathrm{in}} = \hat d$ is valid. The total energy flow is written as 

\begin{equation}
\label{formel19}
\vec{S}=
\underbrace{\frac{c}{8 \pi} \mathfrak{Re}\{\vec{E}_{\mathrm{in}}\times \vec{B}_{\mathrm{in}}^* \}}_{\vec{S}_{\mathrm{in}}} +
\underbrace{\frac{c}{8 \pi} \mathfrak{Re}\{\vec{E}_{\mathrm{in}}\times \vec{B}_{\mathrm{scat}}^* + \vec{E}_{\mathrm{scat}}\times \vec{B}_{\mathrm{in}}^* \}}_{\vec{S}_{\mathrm{ext}}} +
\underbrace{\frac{c}{8 \pi} \mathfrak{Re}\{\vec{E}_{\mathrm{scat}} \times \vec{B}_{\mathrm{scat}}^*\}}_{\vec{S}_{\mathrm{scat}} } \;.
\end{equation}

\noindent
If one compares the situation we investigate with Ref.~\cite{zumofen_prl_2008}, it is important to notice that we have no Gouy phase shift of 90\degree , which is only present in an optical focus. As a consequence, we have no equivalent effect of back reflection as introduced there. This additional phase would lead to a ``perfect reflection'' in the backward direction, respectively to full attenuation in the forward direction~\cite{zumofen_prl_2008}.

We now separate the space into two half spheres, one in the forward direction and one in the backward direction. As shown in Fig.~\ref{fig:01}a, we find that the sum of the energy contributions traveling through the sphere in the forward direction (2) and in the backward direction (1) yield to the same amount with different signs so that their total sum is zero as required by energy conservation. The different energy contributions and their signs, i.e.\ if they are directed inwards to (-) or outwards from the sphere (+), are summarized in the table Fig.~\ref{fig:01}b. In this listing we consider the different parts of the total energy flow. The extinction terms are here listed as negative, since the extinguished power is commonly defined by 
\begin{equation}
P_{\mathrm{ext}} = - \int_{\Omega_{\mathrm{tot}}} \vec{S}_{\mathrm{ext}} \; \mathrm{d} \Omega\;.
\end{equation}

\begin{figure}[t]
  \includegraphics[width=\columnwidth]{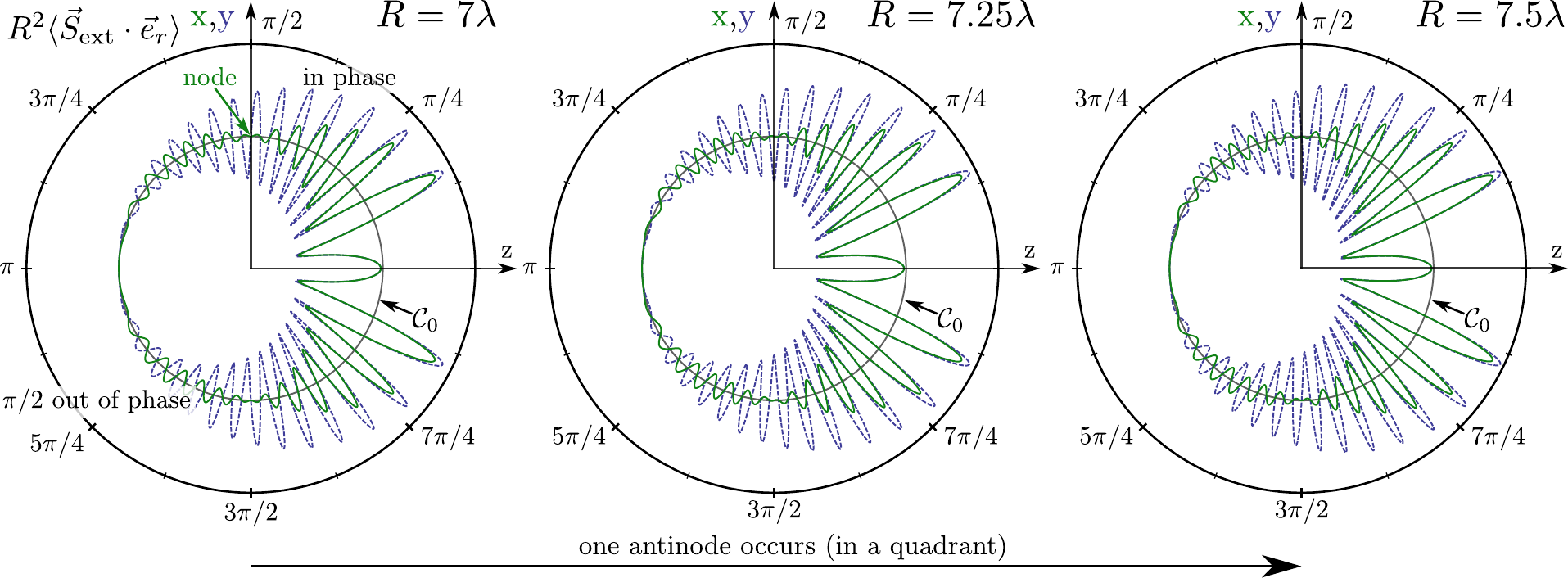}
  \caption{Solutions for $\left.\langle \vec{S}_{\mathrm{ext}}\rangle\right|_{\Theta=\pi/2}$ (blue, dashed) and $\left.\langle \vec{S}_{\mathrm{ext}}\rangle\right|_{\phi=0}$ (green) for values of $R$ chosen in distances of $\lambda/4$. One anti-node occurs within half a wavelength. This periodic behavior has been described in context of Fig.~\ref{fig:01}c. The plots shown here are closely related to the results of Ref.~\cite{Berg:08}.}
  \label{fig:02}
\end{figure}

\noindent
Therefore, the negative signs in the list shown in Fig.~\ref{fig:01}b indicate, that the cross terms yield to an energy flow towards the dipole in the forward direction (2).
The contributions of the incident wave are obvious and equivalent (resp.\ with different signs) for the two half spheres. The scattered energy is greater than zero and has the same amount in both directions i.e.\ the same amount of power is scattered in the forward and backward direction as expected (dipole radiation). The contributions which may not be directly intuitive are the extinction terms. The results of the integration over $\vec{S}_{\mathrm{ext}}$ is shown in Fig.~\ref{fig:01}c. It might be confusing that this integration is not independent of $R$ (radius of the sphere). The incident plane wave has no constant phase on the surface of a half sphere, thus the integral depends on the radius and therefore on how many wave peaks and valleys are collected. This is the reason that Fig.~\ref{fig:01}c shows a periodic behavior in $\lambda$. Moreover, Fig.~\ref{fig:01}c shows, that the sum of the integration over the extinction terms yields always to $-P_{\mathrm{scat}}$ as claimed. Neglecting the oscillations caused by the incident wave it becomes clear that extinction occurs only in the forward direction and the amount adds up to $P_{\mathrm{scat}}$ (c.f.\ Fig.~\ref{fig:03}c), which corresponds to the definition of extinction as scattering plus absorption~\cite{davis_josaa_2001,Berg:08}, since we have no absorption (such as a transfer to heat). Fig.~\ref{fig:01}d shows the extinction part of the Poynting vector in the $xz$- as well as in the $yz$-plane. We have a small constant effect on the $z$-axis in the forward direction, independent of the radius (c.f.\ Fig.~\ref{fig:02}). This can be seen as a small deviation of the reference level $\mathcal{C}_0$ and $R^2\langle \vec{S}_{\mathrm{ext}}\cdot\vec{e}_{\mathrm{r}}\rangle$. This effect occurs, since the phase relation between the scattered and incident wave is constant on the $z$-axis. If the behavior of this plot is studied dependent on the radius ($R$), one anti-node occurs within a half wavelength as shown in Fig.~\ref{fig:02}. Moreover, there is a phase shift of $\pi/2$ between the $\left.\langle \vec{S}_{\mathrm{ext}}\rangle\right|_{\Theta =\pi/2}$ (blue, dashed) and the $\langle \left.\vec{S}_{\mathrm{ext}}\rangle\right|_{\phi =0}$ (green) solutions on the incident half-sphere (1). The sum of the two parts in the forward and the backward direction always yield to zero as expected by energy conservation. 


Since extinction is given by the sum of absorption and scattering it is computed by the total amount of energy extracted from the incident field by the emitter.
We consider the time averaged power extracted by the dipole by~\cite{tretyakov_p_2014} 

\begin{equation}
\label{formel3}
P_{\mathrm{ext}} = -\frac{\omega}{2} \mathfrak{Re}\left\{\int_{V}\vec{J}^*\vec{E}_{\mathrm{in}}\; \mathrm{d}V\right\} = \frac{\omega}{2}\mathfrak{Im}\{\alpha\}\left|\vec{E}_{\mathrm{in}}^0\right|^2 \;. 
\end{equation}

\noindent
Here, $\vec{J}$ is the electric current density (a harmonic time dependence $\exp(-\rm i \omega t)$ was used). For Eqn.~\ref{formel3} it is essential that the electric field is uniform and time harmonic over the volume, $V$, of the system (semi-classical approach). The time averaged energy of the incident wave is 

\begin{equation}\label{formel4}
P_{\mathrm{in}} = \frac{c}{8 \pi} \left|E_{\mathrm{in}}^0\right|^2 \;.
\end{equation}

\noindent
An oscillating dipole radiates power and the the scattered energy is given by~\cite{Jackson1975}

\begin{equation}\label{formel5}
P_{\mathrm{scat}} = \frac{ck^4}{3}|d|^2 = \frac{ck^4}{3}|\alpha|^2 \left|{E _{\mathrm{in}}^0}\right|^2 \;. 
\end{equation}

\noindent
If no energy is transferred into other channels (e.g.\ heat) the amount of extracted and scattered energy has to be equal. This implies that the absorption is zero. In the time averaged case no energy flows into the dipole. All energy which is transferred to the emitter first excites the emitter and is then re-radiated.

\subsection{Introduction of a loss channel, $\beta$}
We now introduce the situation where the emitter is coupled to another bath that can dissipate some of the absorbed power from the incident field. This diminishes the re-radiation of energy and the oscillator is a sink for the energy. The loss contribution of the polarizability can be divided into two parts: One due to scattering loss, described by~\cite{tretyakov_p_2014}

\begin{equation}\label{formel6}
\mathfrak{Im}\left\{\frac{1}{\alpha}\right\} = \frac{2k^3}{3} \;,
\end{equation}

\noindent
and one due to absorption of energy, i.e.\ additional decay of the system, e.g.\ into heat. 
The total inverse polarizability can be generally written as 

\begin{equation}\label{formel7} 
\frac{1}{\alpha} = \xi^{'} - \mathrm{i} \left(\xi ^{''}+\frac{2k^3}{3}\right) \;.
\end{equation}

\noindent
In this notation $\xi^{'}=0$ corresponds to a resonant excitation and $\xi^{'} \to \infty$ implies no polarizability at all. The last term in Eqn.~\ref{formel7} corresponds to the scattering loss factor and $\xi^{''}>0$ refers to the absorption coefficient of the system. For $\xi^{''}=0$ we have a loss-less system i.e.\ the whole extracted energy is re-radiated by the dipole as introduced before. In the following, we want to have a more detailed look at the resonant case ($\xi^{'}=0$). The ideal case, i.e.\ no absorption and under resonant excitation, should be used as a reference. Therefore, we write the polarizability as

\begin{equation}\label{formel8}
\alpha = \frac{\rm i}{\xi^{''} + 2k^3/3} \quad \mathrm{and} \quad \alpha_0 = \frac{3}{2k^3}\rm i \;.
\end{equation}

\noindent
This allows the introduction of a loss parameter, $\beta$, which is sometimes also introduced as a ``single-scattering albedo'',

\begin{equation}\label{formel10}
\beta \equiv \frac{\left|\alpha\right|}{\left|\alpha_0\right|} = \frac{1}{|\alpha_0| \xi^{''} +1} \quad \mathrm{where} \quad 0 \le \beta \le 1 \;.
\end{equation}

\noindent
The dipole moment can be written as

\begin{equation}\label{formel9}
\vec{d} = \alpha \vec{E}_{\mathrm{in}}^{0} = \alpha_0 \beta \vec{E}_{\mathrm{in}}^0 \;.
\end{equation}

\begin{figure}[bh]
  \includegraphics[width=\columnwidth]{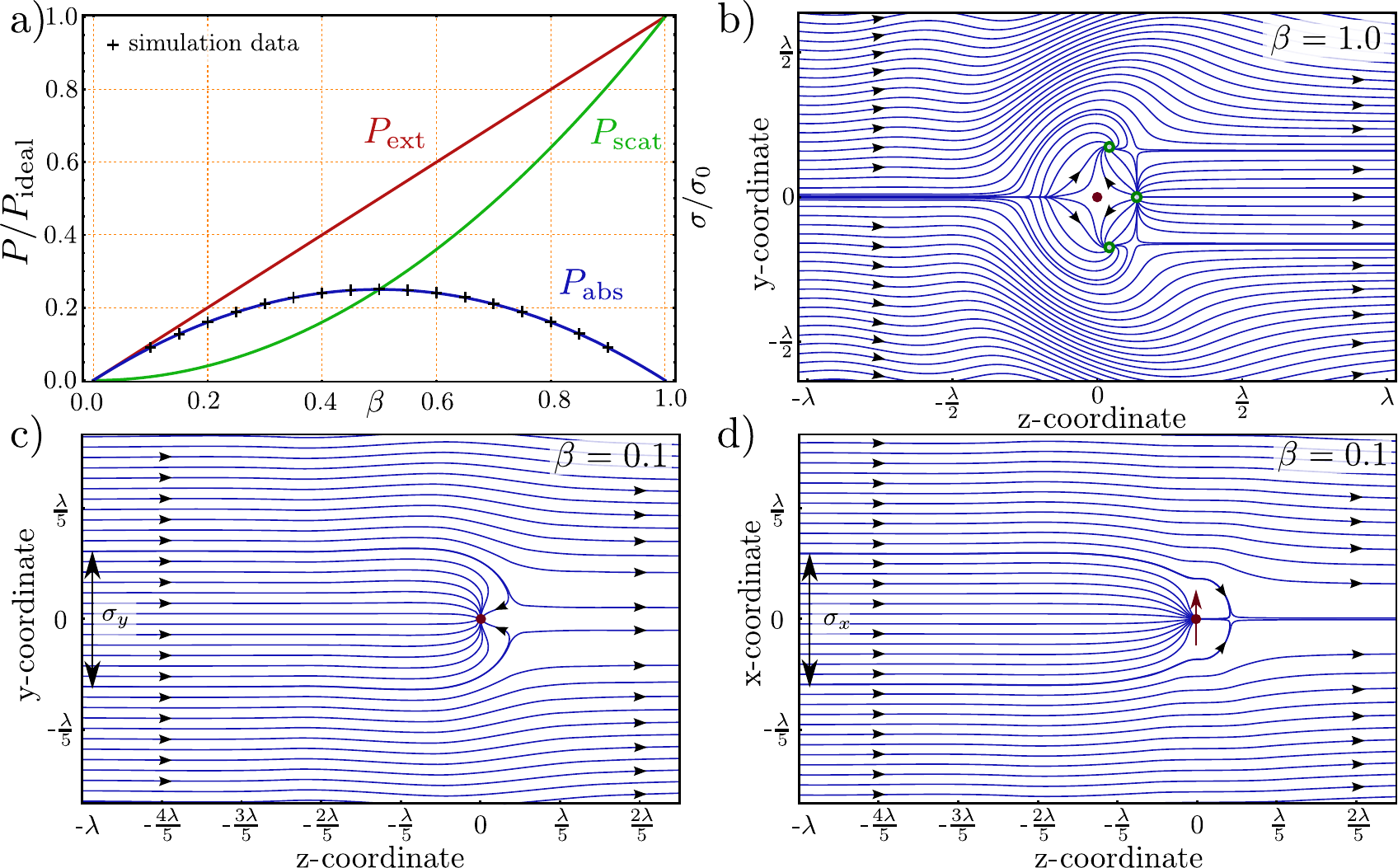}
  \caption{a) Ratio between extinction, scattering and absorption, dependent on the loss parameter $\beta$. At $\beta=0.5$ the absorption, which is an energy transfer to other than coherent re-radiation, displays a maximum. A broadband lossy particle, with $\beta=0$, which has only loss, does not absorb or scatter any light. This is equivalent to a polarizability $\alpha=0$. A perfect scatterer, with $\beta=1$, does not absorb any light. $P_{\mathrm{ideal}}$ corresponds to the case $\beta=1$ and amounts to $(3c)/(4k^2) \left|E_{\mathrm{in}}^0\right|^2$. This is equivalent to the consideration of the cross sections in Eqn.~\ref{formel14}. Due to energy conservation the whole extracted energy is re-radiated. b) Streamlines of the time averaged energy flow of a perfect scatterer in a plane wave. Energy used for its excitation is re-radiated, such that on average no energy is absorbed i.e.\ in the time averaged picture we do not see streamlines ending up in the dipole. c) Streamlines for a lossy system, the loss-parameter $\beta$ is set to 0.1, the emitter is an effective sink of energy. Therefore, some streamlines end up in the emitter. Here, the $yz$-plane is shown. d) same physical situation pictured in the $xz$-plane. The extensions of the cross-section $\sigma_{x}$ and $\sigma_{y}$ are not necessarily the same (see also Fig.~\ref{fig:04}b).}
  \label{fig:03}
\end{figure}

\noindent
In this notation the factor $\beta$ can be seen as a heuristically introduced absorption parameter, since the polarizability of the system is reduced if absorption is introduced. The different energy contributions on resonance are given by

\begin{subequations}
\begin{eqnarray} \label{formel11}
P_{\mathrm{ext}} &=& \frac{3c}{4k^2}\beta |E_{\mathrm{in}}^0|^2 \\ 
P_{\mathrm{scat}} &=& \frac{3c}{4k^2}\beta^2 |E_{\mathrm{in}}^0|^2 \\
P_{\mathrm{abs}} &=& P_{\mathrm{ext}} - P_{\mathrm{scat}} = \frac{3c}{4k^2}(\beta-\beta^2)|E_{\mathrm{in}}^0|^2 \;,
\end{eqnarray}
\end{subequations}

\noindent
where the relation $k=\omega/c$ was used. The three different contributions in Eqn.~\ref{formel19} correspond to the different energy parts given in Eqns.~\ref{formel11},b,c via an integration over an imaginary sphere $\Omega_{\mathrm{tot}}$ (c.f.\ Fig.~\ref{fig:01}) which includes the dipole. \\
Since $\beta$ is given by the ratio 

\begin{equation}\label{formel12}
\beta =\frac{P_{\mathrm{scat}}}{P_{\mathrm{ext}}} = \frac{P_{\mathrm{scat}}}{P_{\mathrm{abs}} + P_{\mathrm{scat}}} \;,
\end{equation}

\noindent
we see that for $\beta<0.5$ the amount of absorbed power dominates the ratio, and for $\beta>0.5$ the amount of scattered energy dominates (c.f.\ Fig.~\ref{fig:03}a). Equation~\ref{formel12} shows the meaning of $\beta$ as the fraction of re-radiated energy to the total extracted energy. In this context it should be mentioned that \textsc{G.\ Wrigge} derives an analogous result for a two-level-system with decay channels (below saturation) in~\cite{Wrigge}, that shows the universality of this fact illustrated in Fig.~\ref{fig:03}a.
It might be confusing that some textbooks (e.g.~\cite{Novotny}) handle Eqn.~\ref{formel3} as the \emph{absorbed} power. If they do so, they use a quasi-static description of the polarizability. In this case scattering by the system is neglected, therefore extinction and absorption become the same. It has to be mentioned that the quasi static polarizability conflicts with the optical theorem, but Eqn.~\ref{formel7} provides a solution to this dilemma. 

To visualize the energy flow we computed the time averaged Poynting vector and solved the differential Eqn.~\ref{formel22} numerically to get its streamlines.

\begin{equation}\label{formel22}
\frac{\mathrm{dr}_{i}}{\rm ds} = S_{i}(x(s),y(s),z(s)) \qquad \mathrm{where} \qquad i \mathrel{\hat=}\{x,y,z\} \;,
\end{equation}
\noindent 
with $s$ used as a ``dummy'' parameter. Such a picture was derived earlier, and allows a vivid insight on the process of light absorption~\cite{paul_spu_1983,bohren_ajop_1983}. 

For the earlier discussed case, where no absorption occurs, we show the case of streamlines in Fig.~\ref{fig:03}b. All streamlines are redirected, but do not end up in the dipole at the origin. It must be noted that the streamlines leave the $xy$-plane. These points are marked with circles in the plot (green). The energy flow for the case of $\beta=0.1$ is shown in Fig.~\ref{fig:03}c,d. It is nicely visible how the dipole collects energy from a region, which is by far larger then the geometrical spread of the dipole (a point-like dipole was considered). Moreover, it is visible that the streamlines are directed towards the dipole, even if they have already passed the position of the emitter. The reason for that behavior is found in the interference terms. The interfere happens in such a way that the energy flow is directed towards the dipole. For a lower $\beta$ value, the effect on the incident plane wave is smaller, as evident on the $xy$-coordinates. With a $\beta$ of unity the distortion of streamlines exceeds one wavelength. This effect is diminished with a lower $\beta$.
\\

Based on the Eqn.~\ref{formel11},b,c we see that $\beta=1$ refers to the ideal system were we have no absorption and the whole energy is re-radiated, whereas the $\beta=0$ case refers to no excitation, respectively no polarizability of the system. Fig.~\ref{fig:03}a shows the normalized energy contributions given in Eqn.~\ref{formel11},b,c plotted as a function of $\beta$. The figure shows nicely the fact that absorption is necessarily concomitant with scattering i.e.\ absorption without scattering can not exist. Moreover, it occurs that the maximum value of absorption is given if $\beta=0.5$, which is equivalent to the same amount of scattered and absorbed energy. As shown in Ref.~\cite{tretyakov_p_2014}, this is also valid for the case of $\xi^{'} \neq 0$, i.e.\ out of resonance.\\

\begin{figure}[bh]
  \includegraphics[width=\columnwidth]{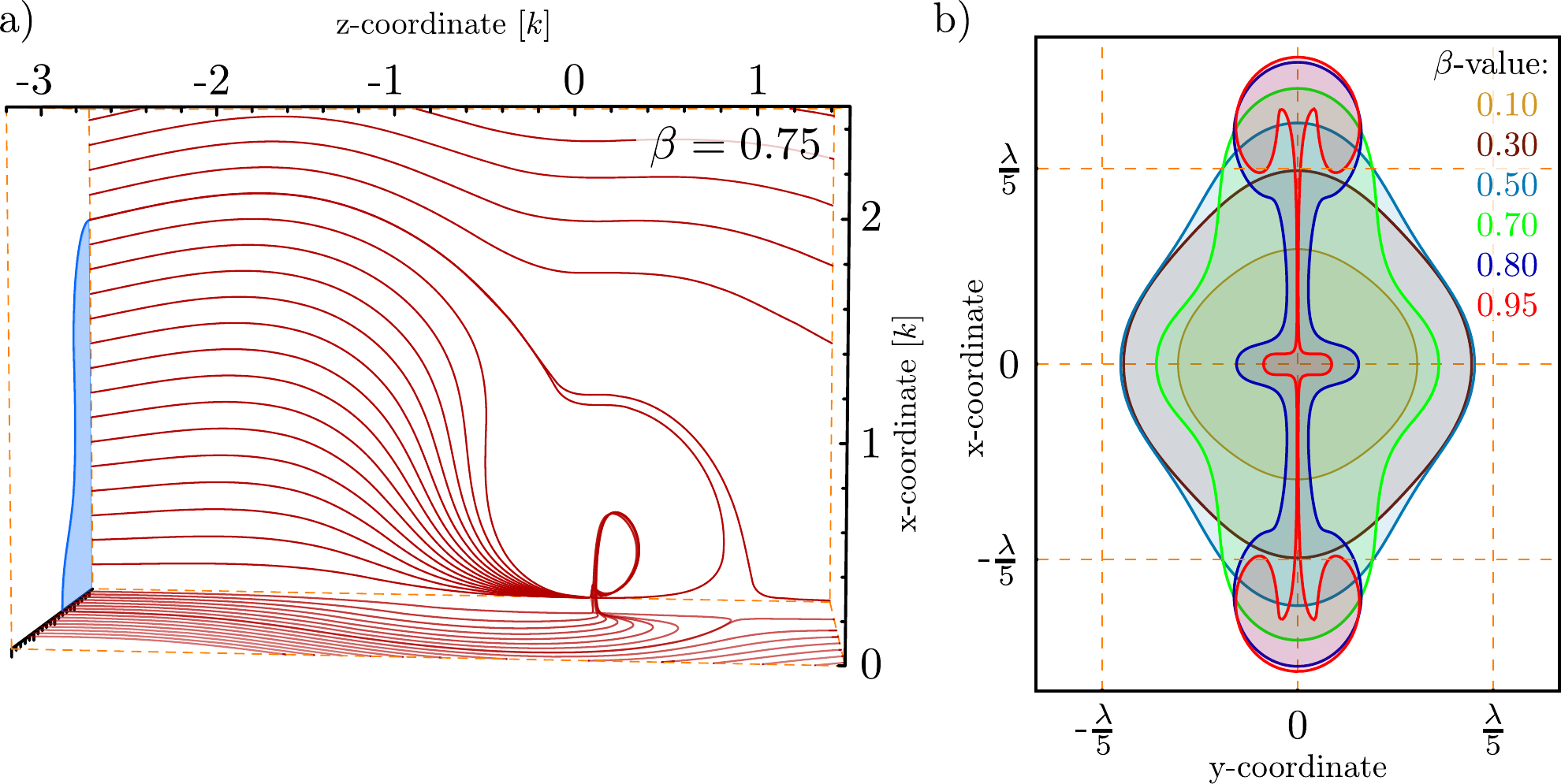}
  \caption{a) Three dimensional representation of the energy flow into a dipole. Due to symmetry, only one quadrant is shown for $\beta=0.75$. It is visible that the streamlines leave the $yz$-plane. By finding the boundary between the lines ending up in the dipole and the ones which are bypassing, we determine the effective absorption aperture. b) the shape of this absorption absorption aperture (equivalent to the cross section) for different values of $\beta$ far away from the emitter ($z$=-15$k$). The crossed labels on the curve $P_{\mathrm{abs}}$ in Fig.~\ref{fig:03}a correspond to the derived areas for the different $\beta$-values in here.}
  \label{fig:04}
\end{figure}

\subsection{The cross sections}
Based on the Eqns.~\ref{formel11},b,c it is possible to define three different on resonant cross sections (it is also possible to find a general description by using Eqn.~\ref{formel7})

\begin{subequations}
\begin{eqnarray}
\sigma_{\mathrm{ext}} &=& \frac{P_{\mathrm{ext}}}{P_{\mathrm{in}}}= \frac{6\pi}{k^2}\beta \\ 
\sigma_{\mathrm{scat}} &=& \frac{P_{\mathrm{scat}}}{P_{\mathrm{in}}} = \frac{6 \pi}{k^2}\beta^2\\
\sigma_{\mathrm{abs}} &=& \sigma_{\mathrm{ext}} - \sigma_{\mathrm{scat}} = \frac{6\pi}{k^2}(\beta-\beta^2)\;.
\end{eqnarray}
\end{subequations}

\noindent
As implied in Fig.~\ref{fig:03}c,d the area of streamlines ending up in the singularity (given by the dipole) suggested a relationship to the absorption cross section. For the three dimensional case, this is depicted in Fig.~\ref{fig:04}a. This definition represents a graphical association to Eqn.~13c. Based on this definition the shape of the absorption cross section was derived for different loss parameters. The result is shown in Fig.~\ref{fig:04}b. To derive the absorption cross section we iteratively searched for the point on the $x$-axis where a small deviation from the point $\vec{r}_{\mathrm{p}}=\{z_0,0,x_{\mathrm{bound}}\}$ yields a change of the streamlines whether they end up in the singularity or not. Once this point is found, the boundary of the aperture can be followed by making circles around the last derived point and checking the endpoint of the streamlines. Such apertures were calculated in the past~\cite{mueller_aeu_1972,shamonina_maapip_2002}. The area of the apertures can than be derived by interpolating the data and using Greens theorem. As visible from the plot (Fig.~\ref{fig:04}b) the shape of the absorption cross section must not necessarily be round.
Moreover, it has to be noted that while the theoretical provided behavior of the absorption cross section as a function of $\beta$ (blue curve in Fig.~\ref{fig:03}a) would intuitively yield an assumption of a shape symmetry against a value of $\beta=0.5$, the shape totally differs between a pair of $\beta$-values symmetric to 0.5 (e.g.\ $\beta=0.3$ and $\beta=0.7$ in Fig.~\ref{fig:04}b). But the computation of the area of the absorption cross section based on the data of the simulation shows a good correspondence to the theoretical forecast (black ``+'' in Fig.~\ref{fig:03}a, i.e.\ the area of the pair $\beta=0.3$ and $\beta=0.7$ is equal). The physical interpretation of this behavior implies that the amount of energy absorbed by a pair of loss parameters symmetric to 0.5 is equal, but the way the energy ``travels'' into the dipole is very different.

For $\beta=1$, i.e.\ no absorption, we find the textbook result for the extinction respectively scattering cross section of a dipole given by
\begin{equation}\label{formel14}
\sigma_0 = \frac{6 \pi}{k^2} = \frac{3 \lambda^2}{2 \pi} \;.
\end{equation}

\noindent
Thus, the whole extinction i.e.\ the reduction of the incident power in the forward direction is due to scattering. 
Leaving the point of an ideal system (without absorption) the absorption cross section is of interest as well. The maximum absorption cross section can be found for a loss factor of 0.5 given by

\begin{equation}\label{formel15}
\sigma_{\mathrm{abs,max}} = \frac{3 \pi}{2 k^3} = \frac{3 \lambda^2}{8 \pi} \;,
\end{equation}

\noindent
again consistent with the result of~\cite{tretyakov_p_2014}. So far, we did not introduce any polarization dependent response of the system, and one should note that the cross sections are indeed independent of the incident wave polarization (i.e.\ linear or circular). The expression for the maximum absorption cross section also holds for the off resonance case~\cite{tretyakov_p_2014}. Thus, a maximum energy dissipation is generally given if $P_{\mathrm{abs}} = P_{\mathrm{scat}}$, which is also visible in Fig.~\ref{fig:03}a.
\noindent
The results which are presented for a Hertzian dipole are equivalent to the analysis of \textsc{E.\ Shamonina} and coworkers, who evaluated the energy flow into a short antenna in~\cite{shamonina_maapip_2002}. The equivalence of the results is not surprising since the inverse polarizability of a short antenna is given by~\cite{tretyakov_p_2014}

\begin{equation}\label{formel23}
\frac{1}{\alpha} = \frac{\rm i \omega}{l^2} (Z_{\mathrm{inp}} + Z_{\mathrm{load}}) \;.
\end{equation}

\noindent
Thus, the math for a short antenna is included in Eqn.~\ref{formel7}. The parameter of interest in Ref.~\cite{shamonina_maapip_2002} is $\Delta$ which corresponds to $\beta$ via $\beta=1/(1 +\Delta)$, i.e.\ the maximal absorption case in this nomenclature is given by $\Delta = 1$. The shape, we found for the absorption cross section of a Hertzian dipole is equivalent to the results of Ref.~\cite{shamonina_maapip_2002}.

\section{Apertures in the Optical Near-field}

Based on the analytically expression of the cross sections given by Eqn.~13a,b,c one can easily disregard, that the definition of the cross section is just strictly valid for $kr \to \infty$. A simple argument to understand this, is that the cross sections are defined via the incident power density $P_{\mathrm{in}}$. Since the total field is given by a superposition of the incident and the scattered field, the power density of the total field differs from the power density of the incident field. Thus, the cross sections given by Eqn.~13a,b,c define an area which yields to the total amount of extinguished, scattered and absorbed power if they are multiplied with $P_{\mathrm{in}}$. Hence, if the energy density of the total field changes the area of the cross section has to change to ensure that the amount of power which is extinguished, scattered or absorbed stays the same. If the total energy density is e.g.\ smaller than the incident power density, the area of the cross sections has to increase, such that energy conservation is fulfilled.

\begin{figure}[t]
  \includegraphics[width=\columnwidth]{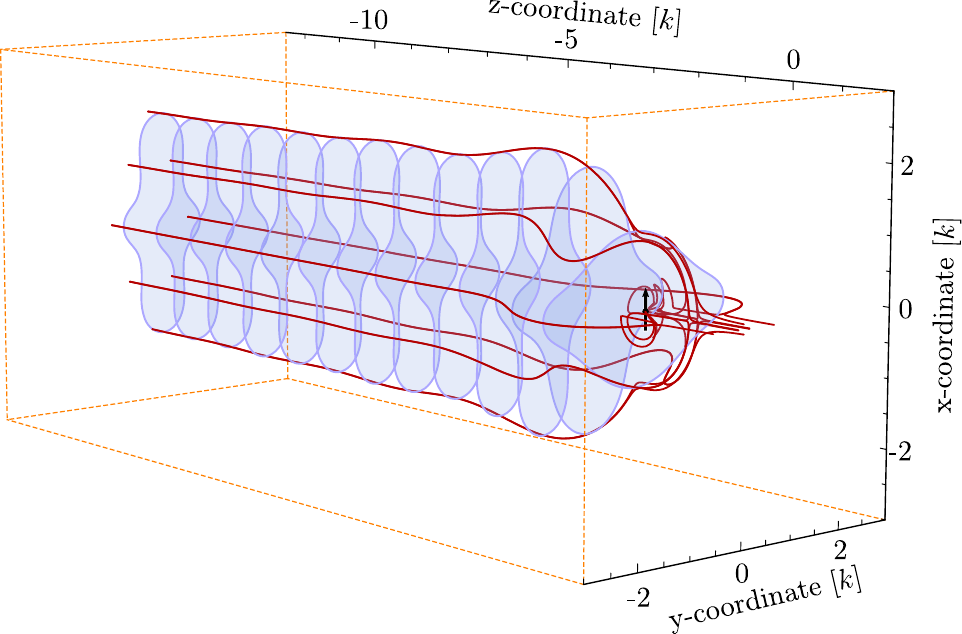}
  \caption{Distance dependence of the absorption cross section. Far from the emitter the streamlines in a defined region end up in the emitter. Approaching the emitter, the same amount of energy is redistributed and the effective cross section enlarges. The energy density inside the cross section is altered. A loss parameter, $\beta$, of 0.75 is assumed, as in Fig.~\ref{fig:04}a.}
  \label{fig:05}
\end{figure}

Fig.~\ref{fig:05} shows the qualitative behavior of the situation described above. The absorption aperture of a dipole with a loss factor of 0.75 is plotted along the way to the dipole, furthermore some boundary solutions on and beside the aperture are shown. At the situation ``far'' away from the dipole, where the power density is given by the incident power density, the shape of the aperture stays constant as expected; approaching the emitter the shape changes. 

\begin{figure}[t]
  \includegraphics[width=\columnwidth]{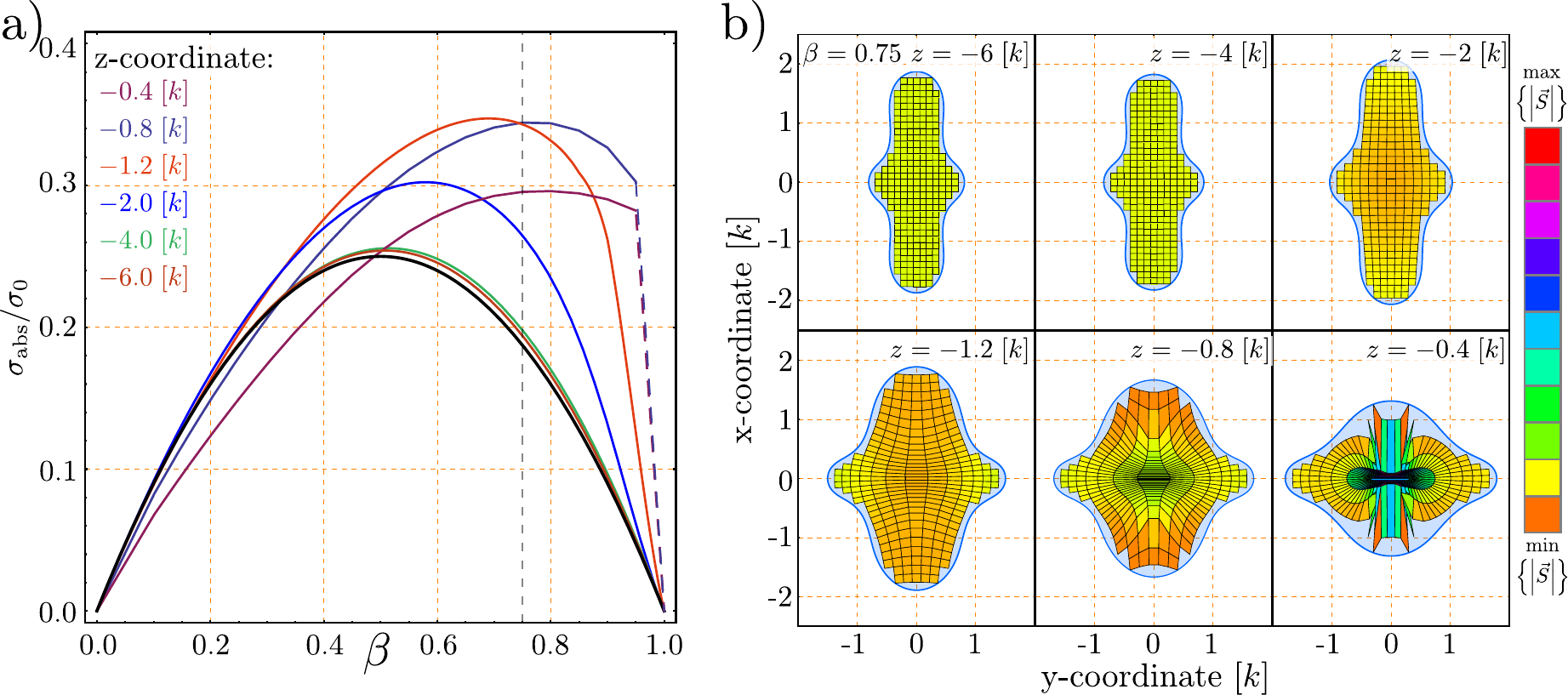}
  \caption{a) Distance dependence of the absorption cross section. Far from the emitter, it corresponds to the theoretical description (black curve). The closer to the emitter the cross section is determined, the peak shifts to higher loss factors $\beta$. The dashed parts of the curves could not be calculated. b) Energy density inside the absorption cross section. We assume a rectangular grid, which is followed along the $z$-axis towards the emitter. It is evident, that the cross section initially widens, and it is then reshaped. As an example the energy density at $k=-1.2$ is obviously reduced, whereas the overall area is enlarged. The corresponding situation in a) is marked with a gray dashed line.}
  \label{fig:06}
\end{figure}

Based on the idea Fig.~\ref{fig:05} provides, a more detailed investigation of the aperture in the optical near-field was performed. The absorption aperture for different loss parameters was evaluated dependent on the distance to the dipole. The results are shown in Fig.~\ref{fig:06}a. The black curve in the plot represents the theoretical curve given by Eqn.~19c. As shown in the plot the area of the aperture increases in the optical near-field. Moreover, it occurs, that there is a maximum value of the aperture for a certain distance in front of the dipole. Following these arguments, this directly yields a change of the total power density. As an example the situation for a loss parameter of $\beta=0.75$ is illustrated in Fig.~\ref{fig:06}b. The density of the energy flow is visualized as a grid of streamlines which were followed on their way towards the dipole. The grid is color-coded corresponding to the energy density of the total field at the position. For the situation in the $xy$-plane at $z$=-6 $[k]$ the energy density appears uniform. For the corresponding situation in Fig.~\ref{fig:06}a we realize, that the area nearly corresponds to the theoretical expectation in the far-field. For the situation where a maximum in the aperture area occurs ($z=$-1.2 $[k]$ c.f.\ Fig.~\ref{fig:06}a), the power density is reduced. Thus, the simulation is conformant to the previous argument.

Fig.~\ref{fig:06}b shows, that the power density does not change homogeneously approaching the emitter. Close to the emitter, the shape of the grid adapts the shape of the dipole radiation pattern (donut-shape). This indicates, that the energy flow in the near-field is dominated by the energy flow of the dipole itself. This corresponds to the result of \textsc{G.\ Zumofen} and coworkers~\cite{zumofen_prl_2008}, when they describe that the dipolar component of an incident focused wave can be perfectly reflected by a single dipole. In our description, the energy flow close to the dipole seems to match such a dipolar pattern. 

\section{Circular Dipole}

\begin{figure}[bh]
  \includegraphics[width=\columnwidth]{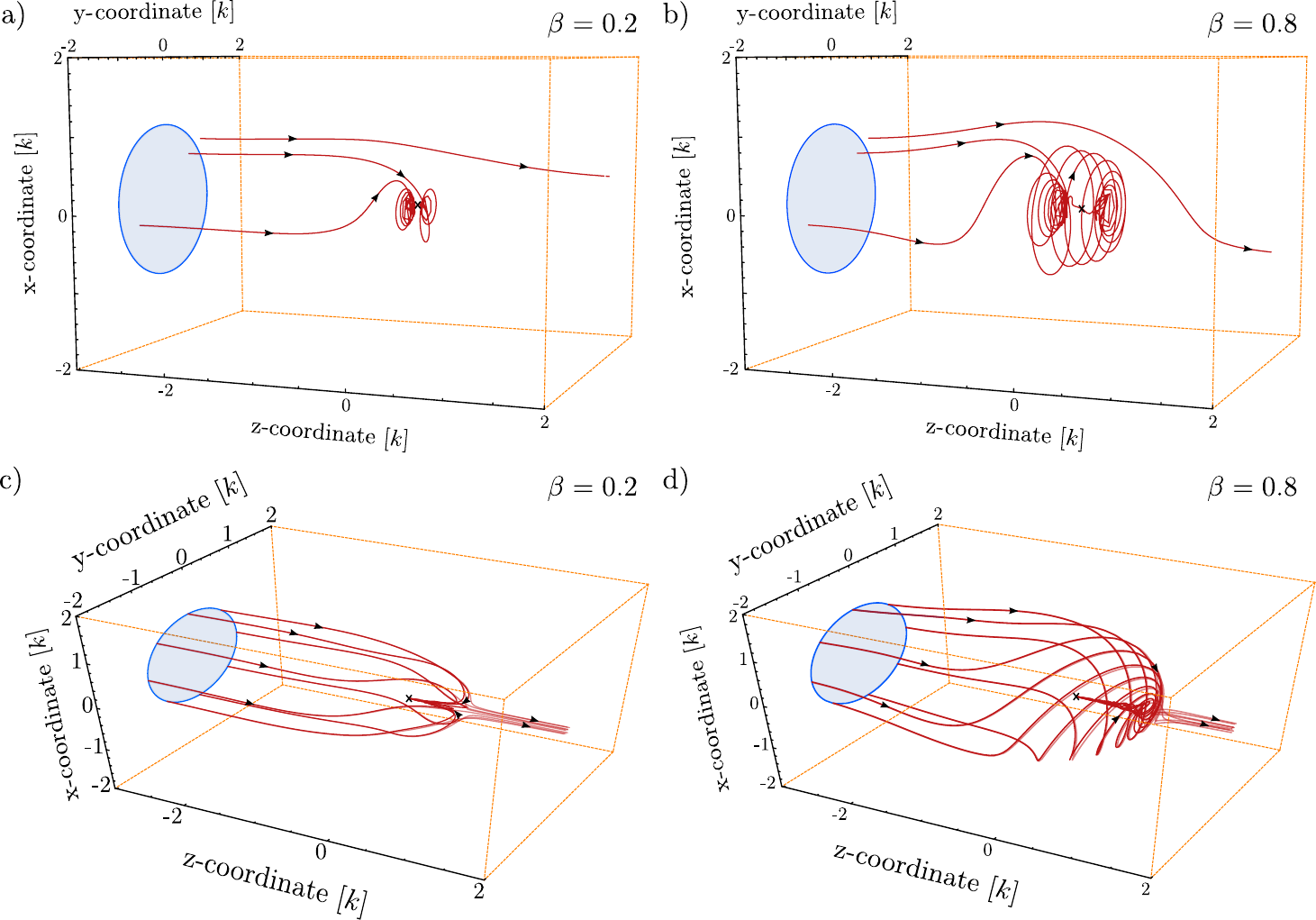}
  \caption{The absorption cross sections for a circular dipole. All incoming aperture shapes are circular. The dipole is located at the origin and marked with an \textbf{x}. The size of the absorption cross section is symmetric around $\beta=0.5$. a) shows the energy flow for $\beta=0.2$ for selected points inside and outside the aperture. The same points are followed in b) for the case of $\beta=0.8$. The size of the cross section is the same. c) and d) show the energy flow for points slightly inside and outside of the aperture. Although the aperture is the same, the energy flow differs in the optical near-field. These pictures correspond to the calculations in~\cite{Arnoldus}, but show the energy flow \emph{into} the dipole.}
  \label{fig:07}
\end{figure}

If one considers an atom as a single emitter it is known from atomic physics that linear as well as circular polarized excitation is of interest. Based on the math introduced in section 2 it is simple to formulate the problem for a circular excitation. With an incident wave given by 

\begin{equation}\label{formel23a}
\vec{E}_{\mathrm{in}} = E_{\mathrm{in}}^0 \hat \epsilon \exp(\rm i k z - \rm i \omega t) \;,
\end{equation}

\noindent
where $\hat \epsilon = 1/\sqrt{2}(1,\rm i ,0)^T$ corresponds to the Jones vector of a circular polarized wave. This yields to a dipole moment (on resonance)

\begin{equation}\label{formel24}
\vec{d} = \alpha_0 \beta \hat \epsilon \exp{(-\rm i \omega t)} = \frac{3}{2\sqrt{2}k^3}\beta(\rm i, -1, 0)^T \exp(-\rm i \omega t) \;. 
\end{equation} 

\noindent
Notice the common phase shift of $\pi/2$ between the incident wave and the reaction of the driven oscillator. The Eqns.~\ref{formel23a} and \ref{formel24} correspond to the Eqns.~\ref{formel2} and \ref{formel9} in section 2. Based on this modification the same investigations as for the Hertzian dipole were performed. Fig.~\ref{fig:07} shows the apertures as well as some streamlines for the circular case and different loss parameters. The way the energy travels into the dipole corresponds to the energy flow of a circular dipole emitter itself (without the incident beam) as investigated in~\cite{Arnoldus}. One has to notice, that the direction of the energy flow in our case is towards the dipole while the energy flow of a radiating dipole is logically outwards. This is due to the interference terms which direct the energy into the dipole. 
If the plots in Fig.~\ref{fig:07}a,b are compared, it is nicely visible that the behavior of the energy flow is more dominated by the scattered part for $\beta>0.5$. If one follows the solution which bypasses the dipole it occurs that the solution first spins around the $z$-axis until it changes the direction and gets sucked up into the dipole. Fig.~\ref{fig:07}c,d shows the boundary solutions for the two cases plotted in Fig.~\ref{fig:07}a,b.
  
\begin{figure}[th]
  \includegraphics[width=\columnwidth]{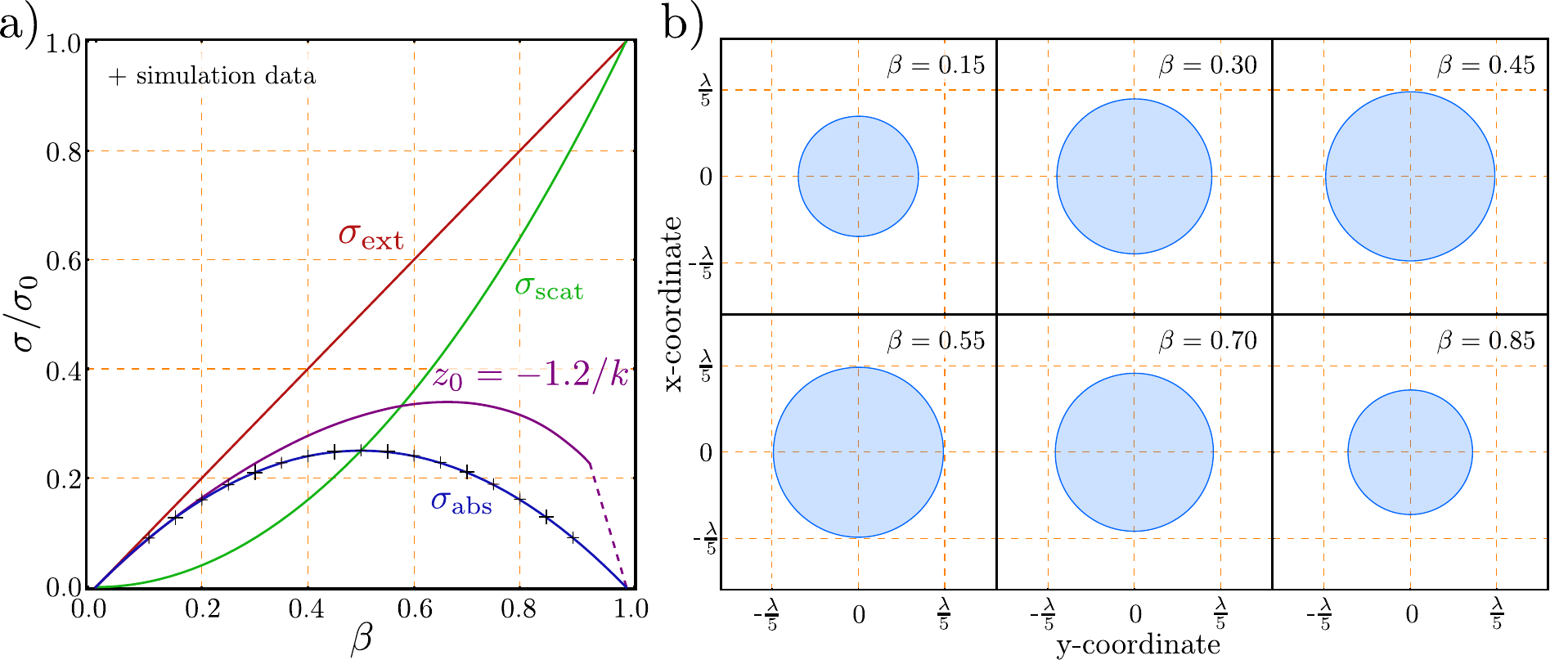}
  \caption{a) The size dependence of a absorption aperture corresponds exactly the case for the Hertzian dipole (see Fig.~\ref{fig:03}a). The cases corresponding to the plots in b) are highlighted with a bold +. b) apertures, far away from the dipolar emitter for different values of the loss parameter $\beta$. The determined values in a) correspond to calculations for different apertures. Unlike for the Hertzian dipole, we observe a shape symmetry against the value of $\beta=0.5$, although the energy flow is different, as shown in Fig.~\ref{fig:07}.}
  \label{fig:08}
\end{figure}

The pair of loss values ($\beta$) chosen for the plots in Fig.~\ref{fig:07} are symmetric to 0.5 considering the theoretical basis (c.f.\ Fig.~\ref{fig:03}a/ Fig.~\ref{fig:08}a) that means the amount of energy absorbed by the dipole is equivalent for both cases. As shown in Fig.~\ref{fig:08}a the simulation data satisfies the theoretical forecast (which is exactly the same as for the linear case) for a circular excitation as well. But if the results are compared to the results of a Hertzian dipole it occurs that there is a shape symmetry against a loss value of 0.5 (Fig.~\ref{fig:08}b). This is more intuitive than for the Hertzian dipole case if one has the symmetric theory plot in mind (c.f.\ Fig.~\ref{fig:03}a). Moreover, it occurs that the shape of the absorption cross section stays perfectly round as visible in Fig.~\ref{fig:08}b. Investigating the energy flow of a symmetric pair of loss values it appears that the way the energy takes differs drastically for both cases (see also Fig.~\ref{fig:07}). 

Comparing the situations for the Hertzian and the circular dipole it is visible that both mathematical descriptions nicely yield a result which satisfies the underlying theory, independent of the incident wave polarization. Whereas the energy flow for the different cases differs totally, but it always fulfills the boundary conditions that the ``total area'' of the absorption cross section yields to a corresponding result. Thus, it might be questionable how meaningful the energy flow itself is. Especially, since the definition of the Poynting vector is not unique. But the results of the simulation provided in the current paper may give an intuitive picture how light absorption occurs.

For the Hertzian dipole, it was determined, that the effective far-field aperture is altered in the optical near-field. This seems to be equivalently the case for the circular dipole. An example is depicted in Fig.~\ref{fig:08}a. These apertures are relatively easy to determine, since their shape is always circular, such that the derivation of one point on the boundary of the aperture is sufficient to calculate the area ($A=\pi r^2$).

\section{Conclusions}

The presented results describe the situation for a plane wave and a dipole emitter, and are fully equivalent for short antennas in the RF- or MW-range. The emitter can be a circular or a linear point-size dipole. To determine the absorption cross-section, the streamlines of the Poynting vector are followed, and result in an effective aperture, which matches the analytical derivations. For a point-like Hertzian dipole, the apertures are not circularly symmetric. The absorption cross-section is fully equivalent in both cases. Whereas for the linear case, the effective aperture changes its shape with the loss parameter, $\beta$, the circular dipole always has a circular shape. 

In both cases, the apertures are changed in the optical near-field. This might not necessarily be interesting for experiments with focused light. Combinations with nano-particles or plasmonic structures might explore the described effects. Extinction experiments in the optical near-field~\cite{gerhardt_prl_2007,Gerhardt2007,Gerhardt} will be likely dominated by other effects, such as the emission of sub-wavelength apertures.

The described math is presently extended to the case of multiple dipolar emitters, which might ``cloak'' each other by their presence. This yields situations comparable to Ref.~\cite{Chen2013}. Furthermore, we extend our approach to the case of a real optical focus, as described earlier~\cite{richards_potrsolsamaps_1959,boivin_josa_1967}. This would correspond to a graphical representation of the case described by Ref.~\cite{zumofen_prl_2008}.

\section*{Acknowledgement}
I.G.\ acknowledges the discussions with G.\ Wrigge and G.\ Zumofen (ETH Z\"{u}rich) regarding this topic. We acknowledge funding from the Max Planck-UBC quantum materials center, the Max Planck Society (J.W., Max Planck fellowship), and the EU via the project SIQS.

\section*{Supplementary information}
The source code (\texttt{Mathematica 9.0.1, Wolfram Research}) to derive Fig.~\ref{fig:03}b,c,d and the effective aperture shape can be found on \href{http://gerhardt.ch/apertures.php}{\texttt{http://gerhardt.ch/apertures.php}}.

\end{document}